\documentclass[12pt]{article}
\usepackage{graphicx}
\title{On the dynamics of  maximum extractable entanglement for open systems}
\author{E. Isasi, D. Mundarain and J. Stephany.
\\
\\
Departamento de F\'{\i}sica, Secci\'{o}n de Fen\'{o}menos \'{O}pticos, \\
Universidad Sim\'{o}n Bol\'{\i}var,\\
Apartado Postal 89000, Caracas 1080A, Venezuela }

\begin{document}

\maketitle
\begin{abstract}
In this work we study the dynamics of the maximum extractable entanglement for a system composed of two qubits interacting either with two  independent thermal  baths,  a common thermal bath or a common squeezed bath. The states with maximum entanglement are found applying   filtering operations  which transform each  state to  a state in Bell diagonal form.  We observe a  revival of the maximum extractable entanglement for  common baths. It is also shown that for some particular states in two independent  baths at zero temperature, one can partially recover the initial entanglement at any time.
\end{abstract}

\section{Introduction}
Entanglement in systems of few particles is one of the most characteristic aspects of quantum dynamics.\cite{EinAPR1935,HorRHH2007}. With its counterpart decoherence is a key element in quantum computing, quantum cryptography and quantum teleportation\cite{BenCBC1993}. Quantification of the degree of entanglement\cite{HorRHH2007,PleMV2005} corresponding to a given  quantum state and  understanding of how it changes due to interactions with other quantum systems  or with the environment are of great importance, both from a fundamental point of view or to envisage possible applications. For systems with many degrees of freedom  recent developments have stressed approximate methods for the separation of the entangled component of a given configuration \cite{HorRHH2007,PleMV2005,LewMS1998}, but a complete characterization has not been achieved. On the contrary for two qubits  the works of   Peres, Horodecki, Hill, and  Wootters \cite{PerA1996,HorMHH1996,HilSW1997,WooW1998}  established the base for a complete discussion of entanglement in terms of  algebraic properties of the density matrix.  In particular, the  Peres-Horodecki criterion \cite{PerA1996,HorMHH1996}  and  concurrence as defined by Wootters  \cite{WooW1998}  allow to quantify entanglement for arbitrary states.

When interacting with the environment, entanglement between subsystems tends to fade away.  Interaction may be represented using a bath with chosen specific properties. In some cases a complete suppression  of entanglement may be observed at finite time, an effect which is referred as entanglement sudden  death. \cite{Eberly}. Since in general,  interaction with some kind of bath is inevitable and one is interested in having maximum entanglement at  disposal it is important to device strategies to preserve or recuperate  entanglement during the evolution of  an open system. Among the strategies  to minimize the influence of the environment are the use of quantum Zeno effect \cite{Maniscalco}, the identification and use of free decoherence spaces \cite{Lidar}, and the application of quantum error correcting codes in quantum computation\cite{Nielsen}.

In this work we are interested in the use of local operations which allow to recuperate at least partially the entanglement which has been  lose in the interaction with the bath. As is known, there are situations where it is possible to improve the degree of entanglement of a pair of systems by means of filtering operations consisting in local non unitary operations and classical communication (LSOCC)\cite{KenLM1999}. Moreover, as we discuss below for each state there exists an optimal filtering operation for which the  image state is the one with maximum entanglement among the accessible sates\cite{KenLM1999}. The concurrence of this state is what is call the maximum extractable entanglement of the original state. In this paper we propose to study the evolution of the maximum extractable entanglement for two qubits in contact with a bath in order to identify conditions for which local observers using (LSOCC) at adequate times may recuperate maximum final entanglement. In the following section we review some of the fundamental concepts discussed in this paper. Then in section (\ref{sec3}) we discuss  the explicit form of the optimal filtering operation. In section (\ref{sec4}) the dynamics of the maximum extractable entanglement  for a system composed of two qubits in the presence of two  independent thermal  baths,  a common thermal bath or a common squeezed bath is studied. In section (\ref{sec5}) the specific case of interaction with the vacuum of two independent baths is discussed in some detail.

\section{Concurrence, filtering operations and entropy}
\label{sec2}
For a composed system with subsystems $A$ and $B$ in  a pure state, a good measure of entanglement is the entropy of either of the two subsystems. The entanglement of formation of a mixed state $\rho$ defined as  the average entanglement of the pure states appearing in its decomposition minimized over all possible decompositions \cite{BenCDS1996} is also a good measure of the entanglement of such system. It is a monotonically increasing function of concurrence introduced in \cite{HilSW1997,WooW1998} which consequently  may be taken as an entanglement measure too. For two qubits, with density matrix $\rho $ , concurrence \cite{WooW1998} is calculated in terms of the eigenvalues $R_1, R_2, R_3,R_4$  of the related
matrix  $R$ defined by,
\begin{equation}
R =  \rho \,\, \sigma_y \, \otimes\, \sigma_y \,\,\rho^{*}\,\,\sigma_y\,\otimes\, \sigma_y .
\end{equation}
It is given by
\begin{equation}
C = {\it max} \{0, 2 \sqrt{R_m}-(\sqrt{R_1}+\sqrt{R_2}+\sqrt{R_3}+\sqrt{R_4})\}\ \  ,
\end{equation}
where
\begin{equation}
R_m = {\it max} \{R_1, R_2, R_3,R_4\} .
\end{equation}
Entanglement  measured, for example, by concurrence diminishes in general by effect of decoherence but may be preserved in some particular situations when decoherence free subspaces are allowed. It may also be partially, but in general not totally, recovered by applying on the system  some specific operations. Among them, filtering operations of the form
\begin{equation}
\label{Filt}
\bar{\rho}=\frac{(A\otimes B)\rho(A\otimes B)^\dagger}{Tr[(A\otimes B)\rho(A\otimes B)^\dagger]}\ \  ,
\end{equation}
are important because they represent  the only way to increase the entanglement of the bipartite system using  local operations and classical communication.  Under the action of these operations  the concurrence transforms as \cite{VerFAD2001},
\begin{equation}
 \label{cprime}
\bar{C}=C\frac{|det(A)||det(B)|}{Tr[A^\dagger A\otimes B^\dagger B \rho]} .
\end{equation}
which in particular  means that separable and entangled states are mapped onto their own kind.

As we mentioned earlier it  is of practical interest to find  the  state with maximum entanglement which can be obtained from the initial state via filtering operations. For two qubits, the density matrix can  be represented in terms of  Pauli matrices ($\sigma_\mu\mapsto\sigma_0=1_{2\times 2},\sigma_i$) as
\begin{equation}
\label{general}
\rho = \frac{1}{4} \left( \sum_{\mu, \nu =0}^3\,  c^{\mu\nu} \,\sigma_\mu^A \otimes \sigma_\nu^B \right) .
\end{equation}
It was shown \cite{KenLM1999} that the optimal filtering operation maps the initial state on a Bell diagonal state. These are states which can be  written in the standard form
\begin{equation}
\label{standard}
\bar{\rho} = \frac{1}{4} \left( 1+ \sum_{i=1}^3\,  C_i \,\sigma_i^A \otimes \sigma_i^B \right) \ \  ,
\end{equation}
and  define a vector sub-space of three dimensions with coordinates $C_i$.  In this subspace, physical states form a tetrahedron  whose vertices $(C_1,C_2,C_3)=\{(-1,1,1),(1,-1,1),(1,1,-1),(-1,1-,1-)\}$ represent pure Bell states. Separable states of this subspace form an octahedron \cite{LeiJMO2006}.
For  matrices written in the standard form  is easy to show that
\begin{equation}
R =  \rho^2 \ \ .
\end{equation}
The eigenvalues of R are $\{\rho_1^2,\rho_2^2,\rho_3^2,\rho_4^2\}$,
and  concurrence in terms of the eigenvalues $\{\rho_1,\rho_2,\rho_3,\rho_4\}$ of $\rho$ is,
\begin{equation}
\label{9}
C = {\it max} \{0, 2 \rho_m-(\rho_1+\rho_2+\rho_3+\rho_4)\}
\end{equation}
where
\begin{equation}
\rho_m  = {\it max} \{\rho_1,\rho_2,\rho_3,\rho_4\}\ \ .
\end{equation}
In the parametrization (\ref{standard}) of the Bell diagonal states the eigenvalues of $\rho$ can be written as
\begin{equation}
\rho_1 = \frac{1}{4}\left( 1+C_1-C_2+C_3 \right)
\end{equation}
\begin{equation}
\rho_ 2=\frac{1}{4}\left( 1-C_1+C_2+C_3 \right)
\end{equation}
\begin{equation}
\rho_3 =\frac{1}{4}\left( 1+C_1+C_2-C_3 \right)
\end{equation}
\begin{equation}
\label{14}
\rho_4=\frac{1}{4}\left( 1-C_1-C_2-C_3 \right)\ \ \ .
\end{equation}

To quantify the degree of mixing of the states during their evolution we use von Neuman entropy defined in terms of the eigenvalues $\rho_i$ of the density matrix as
\begin{equation}
\label{entropy}
S=\sum_i \rho_i\ln \rho_i
\end{equation}
For pure states $S=0$.

\section{Optimal filtering operation}
\label{sec3}

The practical problem we have to address  is given an initial state, to  find explicitly at any time the maximum entangled state and  the optimal filtering operation leading to it. The concurrence of this optimal  state is call the Maximum Extractable Entanglement of the  state at this time. There are different approaches to compute it which depend on the rank of the initial density matrix \cite{VerFDD2001,BenBPS1996,CenWYA2002}.  Following  Leinaas {\it et al} \cite{LeiJMO2006} we  discuss an explicit procedure to perform this mapping, for a five parameter family of states and show how the maximum extractable entanglement evolves in the presence of either two  independent thermal  baths,  a common thermal bath or a common squeezed bath. Verstraete {\it et al} \cite{VerFDD2001} showed that filtering operations (\ref{Filt}) on two qubits correspond to Lorentz transformations
\begin{equation}
 \bar{c}^{\mu\nu} = L^{\mu}_{A \rho} L^{\nu}_{B \sigma} c^{\rho \sigma}
\end{equation}
on the real parametrization (\ref{general}) of the density matrix. The Lorentz transformations $L_A$ and $L_B$ are related with $A$ and $B$ in  (\ref{Filt}) by
\begin{equation}
 L_A = T (A\otimes A^{*}) T^{\dagger} / |det(A)|\ \ \ \ ,\ \ \
 L_B = T (B\otimes B^{*}) T^{\dagger} / |det(B)|
\end{equation}
with $T$ the fixed matrix
\begin{equation}
  T = \frac{1}{\sqrt{2}} \left(
\begin{array}{cccc}
1&0&0&1\\
0&1&1&0\\
0&i&-i&0\\
1&0&0&-1
\end{array} \right) \ \ \ .
\end{equation}
To identify the optimal filtering operations one should first find \cite{LeiJMO2006} the four-vectors $\bar{m}$ and $\bar{n}$ that minimize the function
\begin{equation}
\label{min}
 F(m,n) = c^{\mu \nu} m_{\mu} n_{\nu} \ \ ,
\end{equation}
assuming the normalization condition $ \bar{m}^{\mu} \bar{m}_{\mu} = \bar{n}^{\nu} \bar{n}_{\nu} =1$ and $\bar{m}_{0}\geq 0 $, $\bar{n}_{0}\geq 0 $. Then the optimal Lorentz transformation is chosen so that
\begin{equation}
L^{0}_{A \mu} = \bar{m}_{\mu}, L^{0}_{B \mu} = \bar{n}_{\mu}\ \ .
\end{equation}
In general this transformation does not map directly the initial state to a state written in the standard form (\ref{standard}). An additional local unitary transformation which does not modify the entanglement of the system is necessary to this end. Nevertheless for the class of states  defined by  $C_{4\times4}$ matrices of the form
\begin{equation}
\label{matrixform}
  C_{4\times4} =  \left(
\begin{array}{cccc}
1&0&0&d\\
0&a&0&0\\
0&0&b&0\\
d&0&0&c
\end{array} \right)
\end{equation}
which we consider in this work,  the optimal filtering operation leads directly to a sate in the standard form. For this class of states  using  the Lagrange's multipliers method the extremal problem (\ref{min}) may be solved to find the Lorentz transformation which diagonalize the real parametrization of the density matrix. This is given by,
\begin{equation}
 L_A = L_B =\left(
\begin{array}{cccc}
\beta&0&0&\alpha\\
0&1&0&0\\
0&0&1&0\\
\alpha&0&0&\beta
\end{array} \right)
\end{equation}
where $\alpha$ satisfies both
\begin{equation}\label{equ1}
 \alpha= \frac{-d (1+2 \alpha^2)}{(1+c)\sqrt{1+\alpha^2}}
\end{equation}
and
\begin{equation}\label{equ2}
 \alpha^2 = -\frac{1}{2}+\frac{1}{2} \sqrt{1-\frac{4 d^2}{4 d ^2-(1+c)^2}}
\end{equation}
with
\begin{equation}
\beta = \sqrt{1+\alpha^2}\ \ \ .
\end{equation}
Since the Lorentz transformations are not unitary the transformed density matrix must be normalized after each step. One then obtains the following non zero entries for the final Bell diagonal state:
\begin{equation}\label{equ3}
 C_1 = \frac{a}{\beta^2+2 \alpha \beta d+\alpha^2 c} \ \ ,
\end{equation}
\begin{equation}\label{equ4}
 C_2 = \frac{b}{\beta^2+2 \alpha \beta d+\alpha^2 c} \ \ ,
\end{equation}
\begin{equation}\label{equ5}
 C_3 = \frac{\alpha^2+2 \alpha \beta d+\beta^2 c}{\beta^2+2 \alpha \beta d+\alpha^2 c}\ \ .
\end{equation}
Using these expressions and  Eqs.(\ref{9}-\ref{14}) one  obtains the concurrence of the optimum state.

\begin{figure}[ht]
\centerline{\includegraphics[scale=1.0]{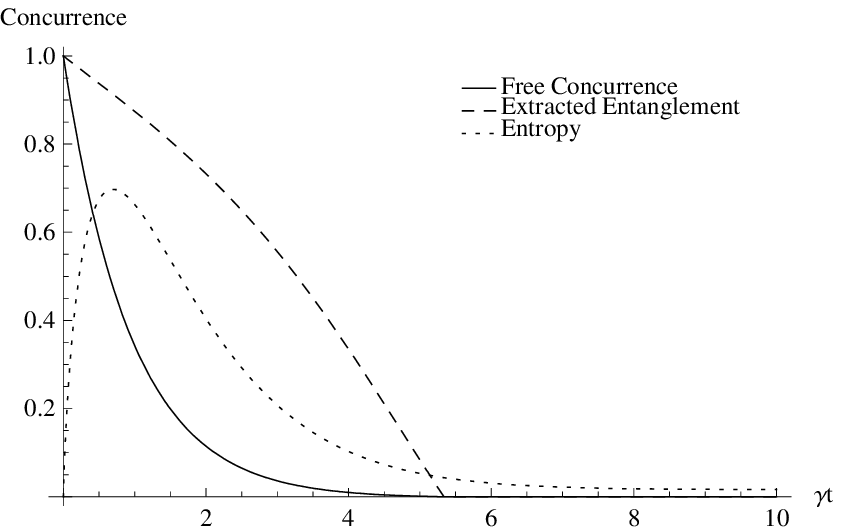}}
\caption{Concurrence and maximum extractable entanglement  evolution for the initial   bell state (1,1,-1) which corresponds to the initial state  $|\Psi(0) \rangle = \frac{1}{\sqrt{2}} \left( |+-\rangle +|-+\rangle \right)$ in  two independent thermal baths, ($n=0.001)$}\label{figure1}
\end{figure}

\begin{figure}[ht]
\centerline{\includegraphics[scale=1.0]{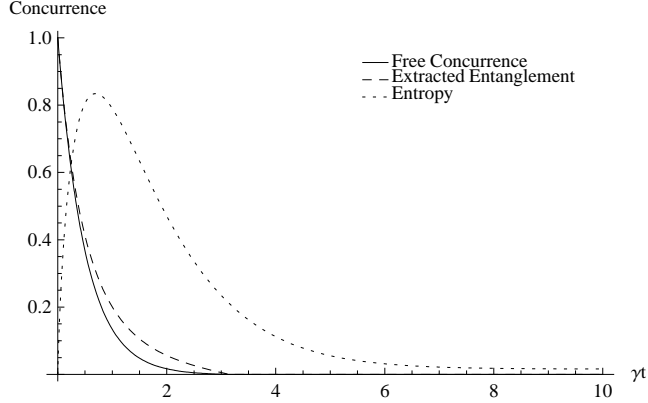}}
\caption{Concurrence and maximum extractable entanglement  evolution for the initial   bell states (1,-1,1) which corresponds to the initial state  $|\Psi(0) \rangle = \frac{1}{\sqrt{2}} \left( |++\rangle +|--\rangle \right)$ in  two independent thermal baths of $n=0.001$}\label{figure2}
\end{figure}

\section{Maximum extractable entanglement evolution}
\label{sec4}

In this section we discuss the evolution of the maximum extractable entanglement  of a two qubits system in three different situations. We consider the two particles interacting with two independent thermal baths, a common thermal bath or a squeezed bath. We note that although the final state in presence of thermal baths is independent of the initial states, the details of the evolution may differ depending of the starting point. Moreover as we discuss in detail below for a squeezed bath the steady state also depends on the initial condition.

Since we are interested in configurations with a high entanglement degree we choose in each case a pure Bell state as the initial configuration.  In what follows we show the results for the initial states $|\Psi(0) \rangle = \frac{1}{\sqrt{2}} \left( |+-\rangle +|-+\rangle \right)$ and  $|\Psi(0) \rangle = \frac{1}{\sqrt{2}} \left( |++\rangle +|--\rangle \right)$. For other possibilities either the initial Bell state belongs to the decoherence free subspace and then the evolution is trivial or  the resulting  evolution is similar to discussed cases.

The master equation for a  pair of two-level particles in the presence of two independent thermal baths is
\begin{eqnarray}
\dot{\rho} &=&\frac{\gamma}{2}\left( (n+1) ( 2 \sigma_a \rho \sigma_a^{\dagger} -\sigma_a^{\dagger}\sigma_a \rho -\rho \sigma_a^{\dagger}\sigma_a )+ n ( 2 \sigma_a^{\dagger}  \rho \sigma_a -\sigma_a \sigma_a^{\dagger}\rho -\rho \sigma_a  \sigma_a^{\dagger}) \right.\nonumber\\
&+&\left. (n+1) ( 2 \sigma_b \rho \sigma_b^{\dagger} -\sigma_b^{\dagger}\sigma_b \rho -\rho \sigma_b^{\dagger}\sigma_b )+ n ( 2 \sigma_b^{\dagger}  \rho \sigma_b-\sigma_b \sigma_b^{\dagger}\rho -\rho \sigma_b  \sigma_b^{\dagger}) \right)
\end{eqnarray}
where $\sigma_a= \sigma_a\otimes1$, $\sigma_b= 1\otimes\sigma_b$ and $\gamma$ is the vacuum decay constant.  We assume  that both baths have the same temperature {\it i.e.}  they have the same average  number $n$ of  thermal photons.

It is easy to show that  evolution with this master equation preserves  the form of states defined by   (\ref{matrixform}). When  Bell states are taken as initial conditions one obtains  for the non zero elements $(a,b,c,d)$ of (\ref{matrixform}) a set of equations whose  solutions  can be found either analytically or numerically. In the particular case $n=0$ the system analytical solutions are discussed  in section (\ref{sec5}).  Meanwhile, using the results of the previous section, we obtain numerically  the evolution of the entanglement and  of the maximum extractable entanglement. In figures  (\ref{figure1}) and (\ref{figure2}) we plot the evolution of concurrence and maximum extractable entanglement for two different initial states. In the first  case we take $|\Psi(0) \rangle = \frac{1}{\sqrt{2}} \left( |+-\rangle +|-+\rangle \right)$ as initial state and in  the second the initial state is  $|\Psi(0) \rangle = \frac{1}{\sqrt{2}} \left( |++\rangle +|--\rangle \right)$.

In this case as in the case with a common bath discussed below sudden death of the entanglement appears sharply. The death time may be computed analytically but the result depends on the initial state and is not particularly illustrating.
In the current case there is a slightly better preservation of entanglement for the  the first initial condition but the entanglement evolution is quite similar for both of them. On the contrary  the evolution of the maximum extractable entanglement  is very different. This difference is particularly evident for early times. It may be understood as the result of a balance between two characteristics of the state: the degree of mixing and the entanglement. For a pure state with non vanishing concurrence the maximum extractable entanglement is always 1. Correspondingly the filtering operations are more efficient on states which are almost pure. On the other hand for a mixed state with a very small concurrence the filtering operation is  not able to improve  the degree of entanglement. This is illustrated with the use of entropy defined in (\ref{entropy}). In figure (\ref{figure1}) the entropy decays sufficiently fast and the maximum extractable entanglement  remains larger until  entanglement sudden death   when of course it also vanish. For the second case (figure (\ref{figure2})) the entropy for the second initial condition decays too slowly and when the state finally is almost pure (entropy almost zero) there is no longer sufficient entanglement to be enhanced.

\begin{figure}[ht]
\centerline{\includegraphics[scale=1.0]{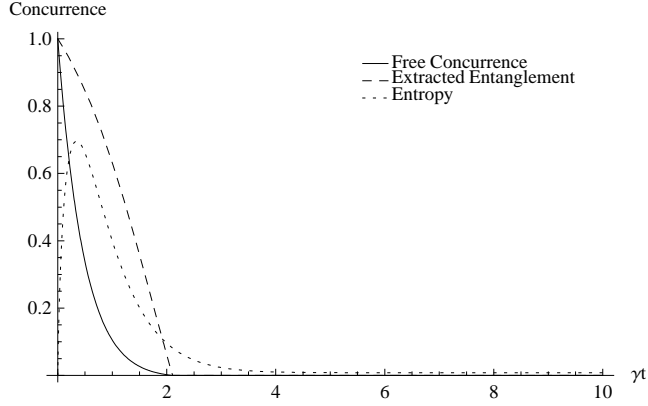}}
\caption{Concurrence and maximum extractable entanglement  evolution for the initial   bell state (1,1,-1) which corresponds to the initial state  $|\Psi(0) \rangle = \frac{1}{\sqrt{2}} \left( |+-\rangle +|-+\rangle \right)$ in  a common thermal bath, ($n=0.001)$}\label{figure3}
\end{figure}

\begin{figure}[ht]
\centerline{\includegraphics[scale=1.0]{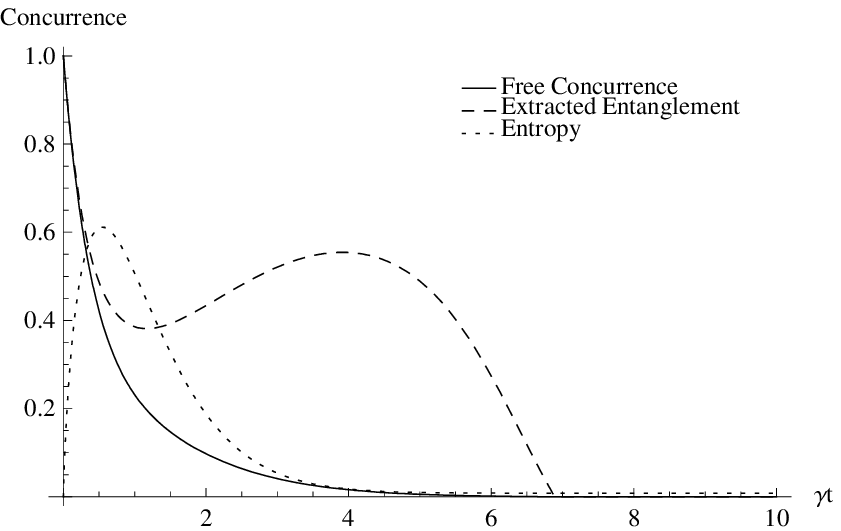}}
\caption{Concurrence and maximum extractable entanglement  evolution for the initial   bell states (1,-1,1) which corresponds to the initial state  $|\Psi(0) \rangle = \frac{1}{\sqrt{2}} \left( |++\rangle +|--\rangle \right)$ in  a common thermal bath of $n=0.001$}\label{figure4}
\end{figure}

For  a pair of two level particles in the presence of a common thermal bath the master equation becomes:
\begin{eqnarray}
\dot{\rho} &=&\frac{\gamma}{2}\left[ (n+1) ( 2 \sigma\rho \sigma^{\dagger} -\sigma^{\dagger}\sigma \rho -\rho \sigma^{\dagger}\sigma)+ n ( 2 \sigma^{\dagger}  \rho \sigma -\sigma \sigma^{\dagger}\rho -\rho \sigma \sigma^{\dagger}) \right]
\end{eqnarray}
where $\sigma= \sigma_a+\sigma_b $. The form of the matrix   (\ref{matrixform}) is also  preserved  in this evolution and  as in the previous case one  obtains the differential equations for to  the non zero elements. In figure (\ref{figure3}) the evolution of concurrence and maximum extractable entanglement is displayed for the initial state $|\Psi(0) \rangle = \frac{1}{\sqrt{2}} \left( |+-\rangle +|-+\rangle \right)$ and in figure  (\ref{figure4}) for the case in which the initial state is $|\Psi(0) \rangle = \frac{1}{\sqrt{2}} \left( |++\rangle +|--\rangle \right)$.

In  figure  (\ref{figure4}) we  observe  that although in this case  concurrence always decreases there is a sector in which  we obtain  a revival of the extractable entanglement.  In other words  a decreasing of concurrence does not always imply  decreasing of maximum extractable entanglement.
This is a more dramatic consequence of the competition between  entanglement and mixing, discussed in the case above.

Finally we consider  evolution of  a pair of two level particles in the presence of a common squeezed bath. The master equation is
\begin{eqnarray}
\label{mastereqsquezz}
\dot{\rho}&=&\frac{\gamma}{2}\left[ (n+1) ( 2 \sigma\rho \sigma^{\dagger} -\sigma^{\dagger}\sigma \rho -\rho \sigma^{\dagger}\sigma)+ n ( 2 \sigma^{\dagger}  \rho \sigma -\sigma \sigma^{\dagger}\rho -\rho \sigma \sigma^{\dagger})\right]\nonumber \\
&-&\frac{\gamma m}{2}\left[e^{i\psi }( 2\sigma ^{\dagger }{\rho }\sigma^{\dagger }-
\sigma ^{\dagger }\sigma ^{\dagger }{\rho }-
{\rho }\sigma^{\dagger }\sigma ^{\dagger })
-e^{-i\psi }( 2\sigma {\rho }\sigma -\sigma \sigma
{\rho }-{\rho }\sigma \sigma)
\right]
\end{eqnarray}
where $m=\sqrt{n(n+1)}$ and $\psi$ are the  parameters of the squeezing.

Once again as in the previous cases, one can verify that the master equation preserves the symmetric form (\ref{matrixform}) of the $C_{4\times4} $ matrix.
\begin{figure}[t]
\centerline{\includegraphics[scale=1.0]{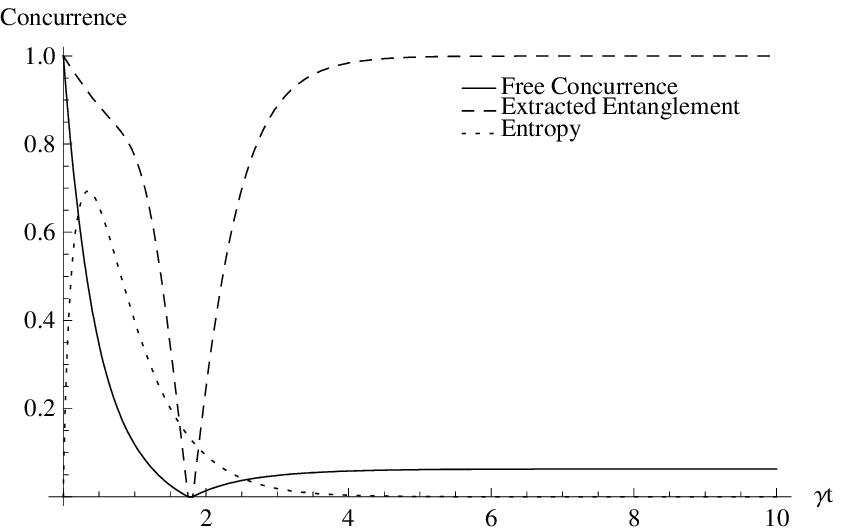}}
\caption{Concurrence and maximum extractable entanglement  evolution for the initial   bell state (1,1,-1) which corresponds to the initial state  $|\Psi(0) \rangle = \frac{1}{\sqrt{2}} \left( |+-\rangle +|-+\rangle \right)$ in  a common squeezed bath, ($n=0.001)$}\label{figure5}
\end{figure}

\begin{figure}[t]
\centerline{\includegraphics[scale=1.0]{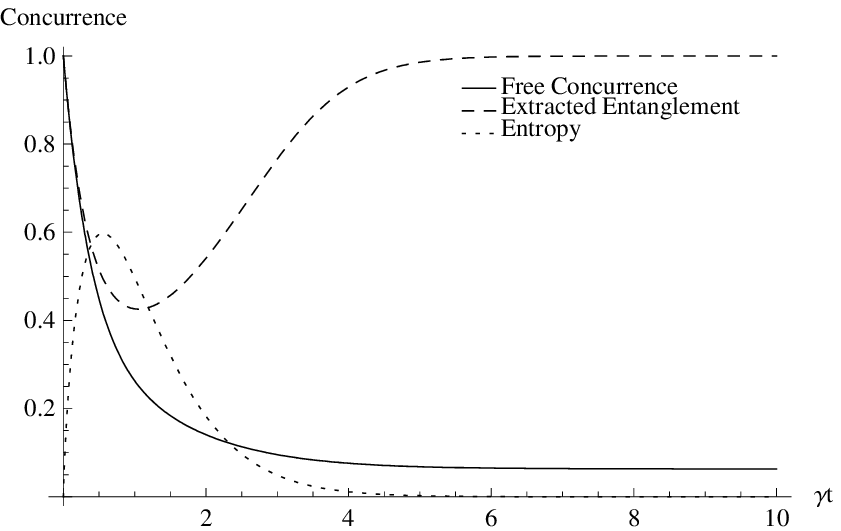}}
\caption{Concurrence and maximum extractable entanglement  evolution for the initial   bell states (1,-1,1) which corresponds to the initial state  $|\Psi(0) \rangle = \frac{1}{\sqrt{2}} \left( |++\rangle +|--\rangle \right)$ in  a common squeezed bath of $n=0.001$}\label{figure6}
\end{figure}

One of the most interesting properties of this system is the existence of  a decoherence free
subspace \cite{Lidar} spanned by the states
\begin{equation}
|\phi _{1}\rangle =\frac{1}{\sqrt{n^{2}+m^{2}}}\left( n|+,+\rangle
+me^{-i\psi }|-,-\rangle \right) ,  \label{l11}
\end{equation}
\begin{equation}
|\phi _{2}\rangle =\frac{1}{\sqrt{2}}\left( |-,+\rangle -|+,-\rangle \right)
,  \label{l12}
\end{equation}
Each of these states is a stationary state of the dynamics defined by the master equation (\ref{mastereqsquezz}) \cite{PalGK1989},\cite{EkeAPB1989},\cite{MunDO2007}. Moreover for states with no contribution of the singlet state $|\phi _{2}\rangle$ the system is driven to the pure state $|\phi _{1}\rangle$. Only for initial conditions with non vanishing components along $|\phi _{2}\rangle$ and some other  direction the final state is a mixed state in the decoherence free subspace\cite{MunDO2007}. This is reflected in the evolution of the system shown in figures (\ref{figure5}) and (\ref{figure6}).
As one can observe in these figures this system displays two interesting effects which are the revival of  entanglement and  of the maximum extractable entanglement. Moreover in the stationary regime almost all the entanglement is recovered. This is explained by the fact that for these initial conditions there is no component in the direction of $|\phi _{2}\rangle$ and explained above the final state is the pure state $|\phi _{1}\rangle$ whose maximum extractable entanglement is 1. This is further illustrated by the behavior of the entropy which in each case goes to zero confirming that the final state is a pure state.

\section{Maximum extractable entanglement in vacuum $n=0$ }
\label{sec5}

\begin{figure}[t]
\centerline{\includegraphics[scale=1.0]{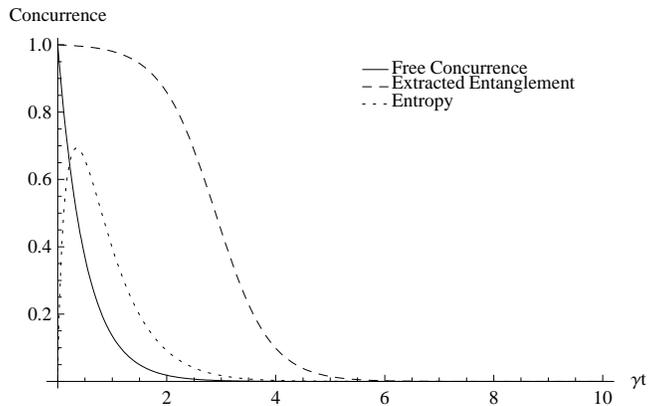}}
\caption{Concurrence and partial extracted entanglement  evolution for the initial   bell states (1,1,-1) which corresponds to the initial state  $|\Psi(0) \rangle = \frac{1}{\sqrt{2}} \left( |+-\rangle +|-+\rangle \right)$ in  a common squeezed bath of $N=0, \alpha = 9$}\label{figure7}
\end{figure}

In this section we  consider the $n=0$ case for two independent baths with  $|\Psi(0) \rangle = \frac{1}{\sqrt{2}} \left( |+-\rangle +|-+\rangle \right)$ as  initial state. The non zero components of the density matrix satisfies the following equations,
\begin{eqnarray}
\dot{d} &= &-\gamma \left( 1+d+2 n d\right)\nonumber\\
\dot{a} &= &-\gamma\left(a +2 n a\right)\nonumber\\
\dot{b} &= &-\gamma \left(b +2 n b\right)\nonumber\\
\dot{c} &= &-2 \gamma\left(d +c +2 n c\right)
\end{eqnarray}
For $n\neq0$ the solutions of these differential equations  are,
\begin{equation}
 d(t) = \frac{e^{-\gamma_0(1+2 n)t}-1}{1+2 n}
\end{equation}
\begin{equation}
a(t)=e^{-\gamma_0 (1+2n )t}
\end{equation}
\begin{equation}
b(t)=e^{-\gamma_0 (1+2n )t}
\end{equation}
\begin{equation}
c(t)= -e^{-2\gamma_0 (1+2 n)t}-\frac{2 e^{-\gamma_0 (1+2 n)t}-e^{-2\gamma_0(1+2n)t}-1}{(1+2n)^2}
\end{equation}
For any finite $ t$ when $n\rightarrow 0$
\begin{equation}
4d^2(t)-(1+c(t))^2 \rightarrow 0
\end{equation}
Then from (\ref{equ1}-\ref{equ2}) and (\ref{equ3}-\ref{equ5}) one obtains
\begin{equation}
\label{infi}
\alpha\rightarrow  \infty \quad C_1 \rightarrow 1 \quad C_2\rightarrow 1\quad C_3\rightarrow -1
\end{equation}
The optimal state at any time is equal to the initial Bell state $(1,1,-1)$.   Using an infinite boost Eq.(\ref{infi}) tells that  the complete initial entanglement is recoverable. One can also partially recover the entanglement to any desired degree at any finite time  using a finite boost. By a numerical analysis one verifies that with the  boost
$\gamma_0 t  \approx \sqrt{\alpha}$ one recovers almost all the entanglement at any finite  time $t$. In figure (\ref{figure7}) we show the extracted entanglement with $\alpha= 9$. In this case one recovers almost all the entanglement if $\gamma_0 t\approx 3$.

\section{Conclusion}
\label{sec6}

In this work we studied the evolution of the maximum extractable entanglement for an open system of two qubits considering three  different interactions with the environment. For two independent thermal baths at zero temperature we show that it is possible to recover almost all the initial entanglement using finite operations of local filtering.  In the case of a common thermal bath we observed during the  evolution an increasing of the maximum extractable entanglement when entanglement was in fact diminishing. Related to this is important to  note that the maximum extractable entanglement is a property of the state and not of the evolution. In this case what is happening is that evolution drove the system to states with less entanglement but more extractable entanglement. This suggest as a strategy to manipulate efficiently the entanglement to set the conditions of interaction of the system with the environment in such a way not to preserve maximum entanglement but maximum extractable entanglement.

\section{Acknowledgments}
This work was supported by Did-Usb Grant Gid-30 and by Fonacit Grant No G-2001000712.

\bigskip

\end{document}